\documentclass[doublecol]{epl2}
\usepackage{amsmath}
\usepackage{amsfonts}
\usepackage{amssymb}
\usepackage{graphicx}
\usepackage{subfig}
\usepackage{bbm}
\usepackage[dvipdfm]{hyperref}

\begin{document}
\title{Renormalization of trace distance and multipartite entanglement close to the quantum phase transitions of one- and two-dimensional spin-chain systems}
\shorttitle{Renormalization of trace distance and multipartite entanglement in 1- and 2-dimensional spin-chains}
\author{Wei Wu \and Jing-Bo Xu \thanks{E-mail: \email{xujb@zju.edu.cn}}}
\institute{Zhejiang Institute of Modern Physics and Department of Physics, Zhejiang University, Hangzhou 310027, People's Republic of China}

\abstract{
We investigate the quantum phase transitions of spin systems in one and two dimensions by employing trace distance and multipartite entanglement along with real-space quantum renormalization group method. As illustration examples, a one-dimensional and a two-dimensional $XY$ models are considered. It is shown that the quantum phase transitions of these spin-chain systems can be revealed by the singular behaviors of the first derivatives of renormalized trace distance and multipartite entanglement in the thermodynamics limit. Moreover, we find the renormalized trace distance and multipartite entanglement obey certain universal exponential-type scaling laws in the vicinity of the quantum critical points.}

\maketitle

The quantum phase transition (QPT) of many-body systems has attracted much attention over the years and has become a hot topic in condensed-matter physics~\cite{1}. The existence of QPT strongly influences the behavior of many-body systems near the quantum critical points associated with the divergence of the correlation length and the vanishing of the gap in the exciton spectrum. How to reveal and characterize the QPT of a many-body system is an important task in condensed-matter physics. Traditional QPT approaches mainly focus on the identification of the order parameters and the pattern of symmetry breaking. Recently, quantum correlation~\cite{2,3,4,5,6,7,8} and trace distance~\cite{9} emerged from quantum information science have been used to characterize QPT. These approaches allow one to detect QPT without any prior knowledge of order parameter.

Trace distance is one of the most fundamental concept in quantum information science and satisfies several attractive properties such as subadditivity, contractivity and strong convexity~\cite{10,11,12}. It is shown that trace distance is an outstanding measure for non-Markovianity in quantum dynamical processes and a witness for initial system-environment correlations in open systems. Moreover, the trace distance can be experimentally obtained using technologies such as quantum-state tomography. Recently, the trace distance $\mathcal{D}(\varrho_{\mathcal{X}\mathcal{Y}},\varrho_{\mathcal{X}}\otimes\varrho_{\mathcal{Y}})$ of the density operator for a composite system $\varrho_{\mathcal{X}\mathcal{Y}}$ and the Kronecker product of its marginals $\varrho_{\mathcal{X}}\otimes\varrho_{\mathcal{Y}}$ has been used to witness the QPTs of coupled cavity lattice systems~\cite{9}. Trace distance approach, which compares a state and its factorized state, is different from the fidelity approach~\cite{add}, which compares two ground states whose Hamiltonian parameters are slightly varied.

On the other hand, quantum entanglement plays a crucial role in quantum mechanics which exhibits the nature of nonlocal correlations in quantum systems and has no classical interpretation~\cite{13}. It is important to find a useful and computable criterion to quantitatively quantify the entanglement in many-body systems~\cite{13}. One of the most widely used measures of multipartite entanglement is residual entanglement which originates from the entanglement monogamy inequality~\cite{14,15}. It has been pointed out that the squared entanglement of formation $E_{f}^{2}$ obeys a general monogamy inequality for arbitrary $N$-qubit states~\cite{15} $E_{f}^{2}(\varrho_{\mathcal{A}_{1}|\mathcal{A}_{2}\mathcal{A}_{3}...\mathcal{A}_{N}})-\sum_{k=2}^{N}E_{f}^{2}(\varrho_{\mathcal{A}_{1}\mathcal{A}_{k}})\geq0$, where $E_{f}(\varrho_{\mathcal{A}_{1}|\mathcal{A}_{2}\mathcal{A}_{3}...\mathcal{A}_{N}})$ quantifies the entanglement between subsystem $\mathcal{A}_{1}$ and subsystem $\mathcal{A}_{2}\mathcal{A}_{3}...\mathcal{A}_{N}$. $E_{f}(\varrho_{\mathcal{A}_{1}\mathcal{A}_{k}})$ denotes the entanglement of formation for the two-qubit system $\mathcal{A}_{1}\mathcal{A}_{k}$. Based on this monogamy property for entanglement of formation, a residual entanglement in terms of squared entanglement of formation is defined by~\cite{15} $\tau(\varrho^{\mathcal{A}_{1}}_{\mathcal{A}_{1}\mathcal{A}_{2}...\mathcal{A}_{N}})=E_{f}^{2}(\varrho_{\mathcal{A}_{1}\mid\mathcal{A}_{2}...\mathcal{A}_{N}})-\sum_{k=2}^{N}E_{f}^{2}(\varrho_{\mathcal{A}_{1}\mathcal{A}_{k}})$. This residual entanglement in terms of squared entanglement of formation can effectively measure the multipartite entanglement which is not stored in pairs of qubits~\cite{14,15}. Moreover, the relevant multiqubit entanglement of formation $E_{f}(\varrho_{\mathcal{A}_{1}\mid\mathcal{A}_{2}...\mathcal{A}_{N}})$ can be analytically calculated. For the case of pure state, i.e., $\varrho_{\mathcal{A}_{1}\mathcal{A}_{2}...\mathcal{A}_{N}}=|\Phi\rangle_{\mathcal{A}_{1}\mathcal{A}_{2}...\mathcal{A}_{N}}\langle\Phi|$, one can obtain~\cite{7,15} $E_{f}(\varrho_{\mathcal{A}_{1}\mid\mathcal{A}_{2}...\mathcal{A}_{N}})=\mathcal{S}(\varrho_{\mathcal{A}_{1}})$, where $\varrho_{\mathcal{A}_{1}}=tr_{\mathcal{A}_{2}...\mathcal{A}_{N}}(|\Phi\rangle_{\mathcal{A}_{1}\mathcal{A}_{2}...\mathcal{A}_{N}}\langle\Phi|)$ and $\mathcal{S}$ is the von Neumann entropy. For the case of mixed state, $E_{f}(\varrho_{\mathcal{A}_{1}\mid\mathcal{A}_{2}...\mathcal{A}_{N}})$ can be obtained via quantum discord~\cite{16,17} by making use of Koashi-Winter formula~\cite{15}.

In this Letter, we study the QPTs of a one- and a two-dimensional anisotropic spin-$1/2$ $XY$ models~\cite{6,8,18} by employing trace distance and multipartite entanglement along with real-space quantum renormalization group (QRG) method which provides a highly efficient approach to treat many-body systems at zero temperature and builds an intuitive and clear physical picture.

We first consider a one-dimensional anisotropic $XY$ spin chain and the Hamiltonian for this system can be described by
\begin{equation}\label{eq:eq1}
H_{XY}=\frac{\lambda}{4}\sum_{j=1}^{\mathcal{N}}[(1+\gamma)\sigma_{j}^{x}\sigma_{j+1}^{x}+(1-\gamma)\sigma_{j}^{y}\sigma_{j+1}^{y}],
\end{equation}
where $\lambda$ denotes the nearest neighbor spin-spin interaction strength, $\gamma$ characterizes the anisotropy of the system and $\sigma^{x,y,z}$ are the usual Pauli matrices.

By making use of the standard QRG approach in Ref.~\cite{4,5,6,7,8}, the original Hamiltonian $H_{XY}$ is decomposed into two parts $H_{XY}=H_{B}+H_{BB}$, where $H_{B}$ and $H_{BB}$ are the block-spin Hamiltonian and the inter-block-spin Hamiltonian, respectively. The explicit expression of $H_{B}$ is given by, $H_{B}=\sum_{l=1}^{\mathcal{N}/3}h_{B,l}$, where $h_{B,l}=\frac{\lambda}{4}[(1+\gamma)(\sigma_{l,1}^{x}\sigma_{l,2}^{x}+\sigma_{l,2}^{x}\sigma_{l,3}^{x})+(1-\gamma)(\sigma_{l,1}^{y}\sigma_{l,2}^{y}+\sigma_{l,2}^{y}\sigma_{l,3}^{y})]$ denotes the $l$-th block-spin Hamiltonian which contains three spins labeled by $1$, $2$ and $3$, respectively. And the inter-block-spin Hamiltonian is given by $H_{BB}=\frac{\lambda}{4}\sum_{l=1}^{\mathcal{N}/3}[(1+\gamma)\sigma_{l,3}^{x}\sigma_{l+1,1}^{x}+(1-\gamma)\sigma_{l,3}^{y}\sigma_{l+1,1}^{y}]$. The degenerate ground states of the $l$-th block-spin Hamiltonian $h_{B,l}$ are
\begin{equation*}
\begin{split}
|\phi_{0}\rangle=&\frac{1}{2\sqrt{1+\gamma^{2}}}(-\sqrt{1+\gamma^{2}}|\uparrow\uparrow\downarrow\rangle+\sqrt{2}|\uparrow\downarrow\uparrow\rangle\\
&-\sqrt{1+\gamma^{2}}|\downarrow\uparrow\uparrow\rangle+\sqrt{2}\gamma|\downarrow\downarrow\downarrow\rangle),
\end{split}
\end{equation*}
\begin{equation*}
\begin{split}
|\phi_{1}\rangle=&\frac{1}{2\sqrt{1+\gamma^{2}}}(-\sqrt{2}\gamma|\uparrow\uparrow\uparrow\rangle+\sqrt{1+\gamma^{2}}|\uparrow\downarrow\downarrow\rangle\\
&-\sqrt{2}|\downarrow\uparrow\downarrow\rangle+\sqrt{1+\gamma^{2}}|\downarrow\downarrow\uparrow\rangle),
\end{split}
\end{equation*}
where $|\uparrow\rangle$ and $|\downarrow\rangle$ are the basis vectors of Pauli matrix $\sigma^{z}$. We construct a projection operator $\mathcal{P}$ to map the original Hamiltonian $H_{XY}$ onto a renormalized Hamiltonian $\tilde{H}_{XY}$ as $\tilde{H}_{XY}=\mathcal{P}^{\dag}H_{XY}\mathcal{P}$ where $\mathcal{P}=\prod_{l}\mathcal{P}_{l}$ and $\mathcal{P}_{l}=|\Uparrow\rangle_{l}\langle\phi_{0}|+|\Downarrow\rangle_{l}\langle\phi_{1}|$. $|\Uparrow\rangle$ and $|\Downarrow\rangle$ are the renamed states for each renormalized spin-$1/2$ particle. Then the effective (renormalized) Hamiltonian of a $XY$ spin chain in one dimension can be expressed as~\cite{6,8}
\begin{equation}\label{eq:eq2}
\tilde{H}_{XY}=\frac{\tilde{\lambda}}{4}\sum_{\ell=1}^{\mathcal{N}/3}[(1+\tilde{\gamma})\sigma_{\ell}^{x}\sigma_{\ell+1}^{x}+(1-\tilde{\gamma})\sigma_{\ell}^{y}\sigma_{\ell+1}^{y}],
\end{equation}
where $\tilde{\lambda}/\lambda=\frac{3\gamma^{2}+1}{2(1+\gamma^{2})}$ and $\tilde{\gamma}=\frac{\gamma^{3}+3\gamma}{3\gamma^{2}+1}$. By solving the QRG equation $\tilde{\gamma}=\gamma$, one can obtain two stable fixed points ($\gamma=\pm1$) and one unstable fixed point ($\gamma=0$) which corresponds to the QPT point of this spin-chain $\gamma_{c}=0$. The QPT point derived from QRG approach is in good agreement with the result obtained by Jordan-Wigner transformation~\cite{1}.

\begin{figure}
\centering
\includegraphics[angle=0,width=8.5cm]{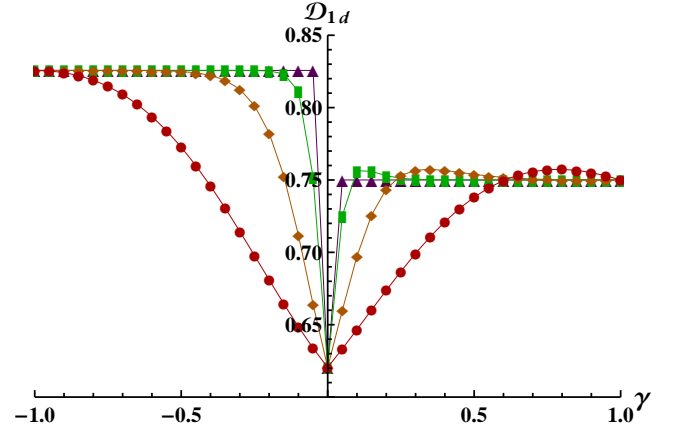}
\caption{\label{fig:fig1} The trace distance $\mathcal{D}_{1d}$ is plotted as the function of $\gamma$ with different QRG iteration steps: 0-th QRG step (red circles), 1-st QRG step (orange diamonds), 2-nd QRG step (green rectangles) and 6-th QRG step (purple triangles).}
\end{figure}
\begin{figure}
\centering
\includegraphics[angle=0,width=8.5cm]{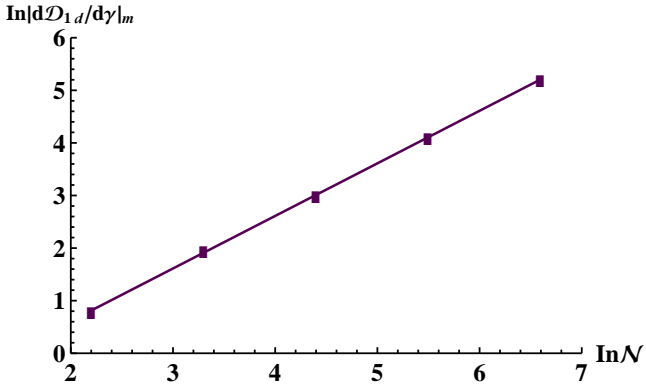}
\caption{\label{fig:fig2} The logarithm of the maximum of $|d\mathcal{D}_{1d}/d\gamma|$ versus $\mathcal{N}$. The purple solid line is obtained by using the least-squares fit of the form: $\ln|d\mathcal{D}_{1d}/d\gamma|_{m}=\theta_{1}\ln\mathcal{N}+c_{1}$ where $\theta_{1}\simeq 0.9994$ and $c_{1}\simeq-1.3861$.}
\end{figure}

Next, we investigate the quantum criticality of the one-dimensional anisotropic $XY$ spin-chain by employing the trace distance along with the QRG method. The definition of trace distance for two operators $\mathcal{O}_{1}$ and $\mathcal{O}_{2}$ is $\mathcal{D}(\mathcal{O}_{1},\mathcal{O}_{2})=\frac{1}{2}\|\mathcal{O}_{1}-\mathcal{O}_{2}\|_{1}$~\cite{9,10,11,12}, where $\|\mathcal{O}\|_{1}$ denotes the trace-norm of any operator $\mathcal{O}$: $\|\mathcal{O}\|_{1}=tr\sqrt{\mathcal{O}^{\dagger}\mathcal{O}}$. If $\mathcal{O}$ is trace class and self-adjoint with eigenvalues $o_{i}$, this formula reduces to the sum of the absolute eigenvalues $\|\mathcal{O}\|_{1}=\sum_{i}|o_{i}|$~\cite{9,10,11,12}. In this Letter, we generalize this method to detect QPTs of one- and two-dimensional spin-chain systems along with QRG. To this aim, we split the $l$-th block-spin Hamiltonian $h_{B,l}$ into two parts, subsystem $\mathcal{X}$ and subsystem $\mathcal{Y}$, i.e., we regard spin $1$ as the subsystem $\mathcal{X}$ and spins $2+3$ as the other subsystem $\mathcal{Y}$. It is necessary to point out that how to choose the subsystems does not influence our physical results, which means we can also choose spins $2+3$ (or $1+3$) as the subsystem $\mathcal{X}$ and spin $2$ (or $1$) as the subsystem $\mathcal{Y}$. Then, the trace distance of the ground state $\varrho_{123}=|\phi_{0}\rangle\langle\phi_{0}|$ and the Kronecker product of its marginal state $\varrho_{1}\otimes\varrho_{23}$ is given by $\mathcal{D}_{1d}\equiv\frac{1}{2}\||\phi_{0}\rangle\langle\phi_{0}|-\varrho_{1}\otimes\varrho_{23}\|_{1}$, where $\varrho_{1}=tr_{23}(|\phi_{0}\rangle\langle\phi_{0}|)$ and $\varrho_{23}=tr_{1}(|\phi_{0}\rangle\langle\phi_{0}|)$. It is easy to check that choosing $\varrho_{123}=|\phi_{1}\rangle\langle\phi_{1}|$ yields the same physical results.

The trace distance $\mathcal{D}_{1d}$ as the function of parameter $\gamma$ for different QRG iteration steps is plotted in Fig.~\ref{fig:fig1}. We find that the trace distance displays a sudden drop near the QPT point $\gamma=\gamma_{c}=0$ which means that the first derivative of trace distance $d\mathcal{D}_{1d}/d\gamma$ is discontinuous at the QPT point as the size of system $\mathcal{N}\rightarrow \infty$. After enough iteration steps, the trace distance $\mathcal{D}_{1d}\simeq0.750$ for $\gamma>\gamma_{c}$, $\mathcal{D}_{1d}\simeq0.825$ for $\gamma<\gamma_{c}$ and $\mathcal{D}_{1d}\simeq0.618$ for $\gamma=\gamma_{c}$.

To further understand the relation between the renormalized trace distance and QPT, we also explore the finite-size scaling behaviors of renormalized trace distance close to the critical points. We find the maximum value of $|d\mathcal{D}_{1d}/d\gamma|$ obeys the following finite-size scaling law: $\ln|d\mathcal{D}_{1d}/d\gamma|_{m}\simeq \theta_{1}\ln\mathcal{N}+const.$ (see Fig.~\ref{fig:fig2}). Numerical analysis tells us that the critical exponent of trace distance $\theta_{1}\simeq1$, this result is in good agreement with previous studies~\cite{6,8} and demonstrates the renormalized trace distance of the density operator for a composite system $\mathcal{XY}$ and the Kronecker product of its marginals $\mathcal{X}\otimes\mathcal{Y}$ can be used to characterize the QPT of $XY$ spin-chain in one dimension.

Here, we would like to briefly discuss why the renormalized trace distance is able to characterize the QPTs of spin-chain systems. The trace distance is a direct measure of how far apart of two quantum states in the state space, the ``difference" between $\varrho_{123}$ and $\varrho_{1}\otimes\varrho_{23}$ consists in correlations of renormalized spin $1$ and spins $2+3$, thus, $\mathcal{D}_{1d}$ can effectively reveal the spin-spin correlations between renormalized spin $1$ and spins $2+3$. As we have described in the QRG approach, the original system $H_{XY}$ can be effectively rescaled to three spins with the renormalized couplings. Thus, after enough iteration steps, renormalized trace distance represents the block-block correlations between two parts of the spin-chain which is strongly influenced by the QPT of spin-chain. In this sense, renormalized trace distance can be used to characterize the QPTs of spin-chain systems.

\begin{figure}
\centering
\includegraphics[angle=0,width=8.5cm]{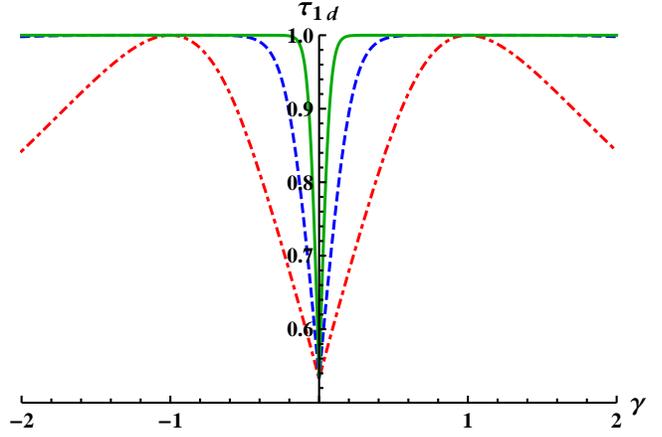}
\caption{\label{fig:fig3} The tripartite entanglement indicator $\tau_{1d}$ is plotted as the function of $\gamma$ with different QRG iteration steps: 0-th QRG step (red dot-dashed line), 1-st QRG step (blue dashed line) and 2-nd QRG step (green solid line).}
\end{figure}
\begin{figure}
\centering
\includegraphics[angle=0,width=8.5cm]{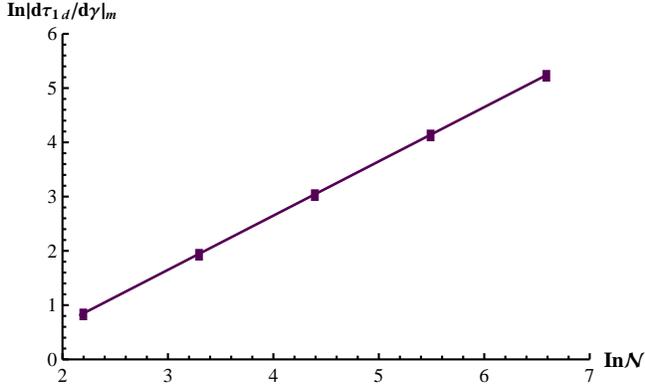}
\caption{\label{fig:fig4} The logarithm of the maximum of $|d\tau_{1d}/d\gamma|$ versus $\ln\mathcal{N}$. The purple solid line is obtained by using the least-squares fit of the form: $\ln|d\tau_{1d}/d\gamma|_{m}=\theta_{2}\ln\mathcal{N}+c_{2}$ where $\theta_{2}\simeq 0.9989$ and $c_{2}\simeq-1.3500$.}
\end{figure}
\begin{figure}
\centering
\includegraphics[angle=0,width=5cm]{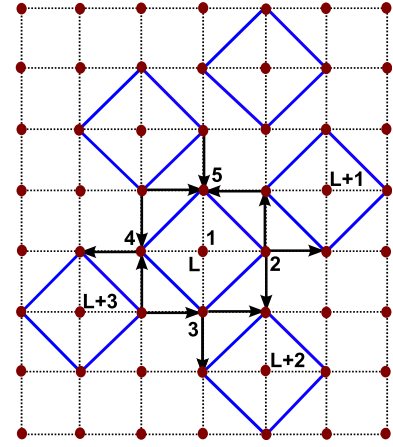}
\caption{\label{fig:fig5} Construction of block-spins with five spins in two-dimensional square lattice. The inter-block interactions are shown by direction of arrows.}
\end{figure}

In the previous discussion, we study the QPT of one-dimensional $XY$ spin-chain by the renormalized trace distance. It is necessary to emphasize that the existence of QPT is independent of the chosen renormalized physical quantity. In order to illustrate this conclusion, we also investigate the QPT of one-dimensional $XY$ spin-chain by the renormalized multipartite entanglement measured by residual entanglement in terms of squared entanglement of formation for density matrix $\varrho_{123}=|\phi_{0}\rangle\langle\phi_{0}|$ (it is easy to check that choosing $|\phi_{1}\rangle$ yields the same physical results). We can obtain a tripartite entanglement indicator $\tau_{1d}(\varrho_{123}^{1})\equiv E_{f}^{2}(\varrho_{1|23})-E_{f}^{2}(\varrho_{12})-E_{f}^{2}(\varrho_{13})$, where $E_{f}(\varrho_{1|23})=\mathcal{S}(\varrho_{1})$ with $\varrho_{1}=tr_{23}(\varrho_{123})$.

The tripartite entanglement indicator $\tau_{1d}$ as the function of parameter $\gamma$ with different QRG iteration steps are displayed in Fig.~\ref{fig:fig3}. We find that the tripartite entanglement $\tau_{1d}$ exhibits a sudden drop near the QPT point $\gamma_{c}=0$ which implies the first derivative of tripartite entanglement indicator $d\tau_{1d}/d\gamma$ is discontinuous at $\gamma_{c}=0$ in the thermodynamics limit. After enough iteration steps, the value of tripartite entanglement remains as constant $\tau_{1d}\simeq 0.532$ for $\gamma=\gamma_{c}$ and $\tau_{1d}\simeq 1$ for $\gamma\neq \gamma_{c}$.

Similar to the case of trace distance, we find the maximum of the first derivative of tripartite entanglement $|d\tau_{1d}/d\gamma|$ exhibits the scaling behavior $\ln|d\tau_{1d}/d\gamma|_{m}\simeq \theta_{2}\ln\mathcal{N}+const.$ (see Fig.~\ref{fig:fig4}). The critical exponents of tripartite entanglement $\theta_{2}$ is very close to 1 which coincides with the critical exponents of trace distance $\theta_{1}$, this result convinces us that the renormalized multipartite entanglement truly captures the quantum critical behaviors of the $XY$ spin-chain in one dimension.

Here, we want to point out that the multipartite entanglement employed in this paper maybe a better indicator of QPT than bipartite entanglement or quantum correlation measures. Most of the existing studies of entanglement in spin-chain models have restricted their attentions to bipartite quantum correlations. However, it has been reported that the bipartite entanglement measured by concurrence may fail to reveal the real quantum critical points of spin-chain systems~\cite{2,19,20}. Moreover, certain important types of entanglement in spin-chain systems (e.g., various $n$-tangles) can not be captured by bipartite entanglement measures. Thus, multipartite entanglement may have some advantages over bipartite entanglement or quantum correlation to reveal QPTs of spin-chain systems~\cite{20,21}. Additionally, the multipartite entanglement measured by residual entanglement has a more concise expression than that of genuine multipartite negativity~\cite{22} which also suggests the multipartite entanglement measured by residual entanglement is a better indicator of QPT. These are the reasons why we choose multipartite entanglement measured by residual entanglement in terms of squared entanglement of formation to reveal QPT.

\begin{figure}
\centering
\includegraphics[angle=0,width=8.5cm]{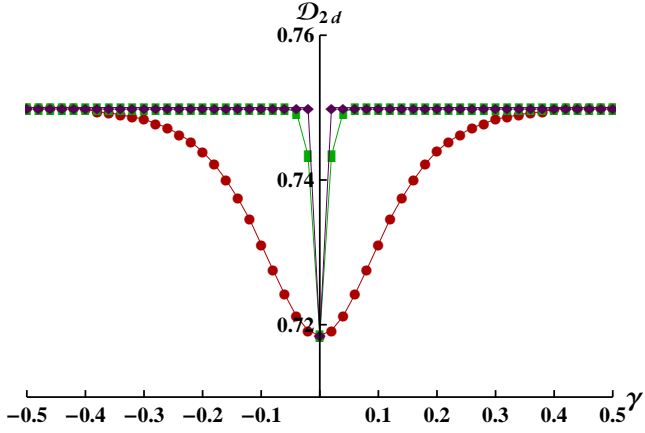}
\caption{\label{fig:fig6} The trace distance $\mathcal{D}_{2d}$ is plotted as the function of $\gamma$ with different QRG iteration steps: 0-th QRG step (red circles), 1-st QRG step (green rectangles) and 2-nd QRG step (purple diamonds).}
\end{figure}

Up to now, we have demonstrated that the renormalized trace distance and multipartite entanglement can be regarded as good signatures of QPTs in one-dimensional spin-chain systems. It is necessary to point out that detecting the QPT by using trace distance and multipartite entanglement along with QRG approach can be generalized to the spin-chain system in two dimensions. The Hamiltonian of a two-dimensional anisotropic spin-$1/2$ $XY$ model represented by the square lattice of $\mathcal{N}\times\mathcal{N}$ spins can be described as follows~\cite{18}
\begin{equation}\label{eq:eq3}
\begin{split}
H_{XY}^{(2d)}=&\frac{\lambda}{4}\sum_{i=1}^{\mathcal{N}}\sum_{j=1}^{\mathcal{N}}[(1+\gamma)(\sigma_{i,j}^{x}\sigma_{i+1,j}^{x}+\sigma_{i,j}^{x}\sigma_{i,j+1}^{x})\\
&+(1-\gamma)(\sigma_{i,j}^{y}\sigma_{i+1,j}^{y}+\sigma_{i,j}^{y}\sigma_{i,j+1}^{y})],
\end{split}
\end{equation}
Similar to the process carried out in the case of one-dimensional $XY$ spin-chain, the total Hamiltonian $H_{XY}^{(2d)}$ is split into two parts as $H^{(2d)}_{XY}=H_{B}^{(2d)}+H_{BB}^{(2d)}$, where $H_{B}^{(2d)}=\sum_{L=1}^{\mathcal{N}/5}H^{(2d)}_{B,L}$ denotes the block-spin Hamiltonian and $H^{(2d)}_{B,L}=\frac{\lambda}{4}\sum_{m=2}^{5}[(1+\gamma)\sigma_{L,1}^{x}\sigma_{L,m}^{x}+(1-\gamma)\sigma_{L,1}^{y}\sigma_{L,m}^{y}]$ is the $L$-th block-spin Hamiltonian, each consisting of five spins (labeled by 1, 2, 3, 4 and 5, respectively), with one spin (1) at the center of the square and four spins (2-5) at the corners (see Fig.~\ref{fig:fig5}). The inter-block-spin Hamiltonian $H_{BB}^{(2d)}$ is given by
\begin{equation*}
\begin{split}
&H^{(2D)}_{BB}=\frac{\lambda}{4}\sum_{L=1}^{\mathcal{N}/5}[(1+\gamma)(\sigma_{L,2}^{x}\sigma_{L+1,3}^{x}+\sigma_{L,2}^{x}\sigma_{L+1,4}^{x}\\
&+\sigma_{L,2}^{x}\sigma_{L+2,5}^{x}+\sigma_{L,3}^{x}\sigma_{L+2,4}^{x}+\sigma_{L,3}^{x}\sigma_{L+2,5}^{x}+\sigma_{L,4}^{x}\sigma_{L+3,5}^{x})\\
&+(1-\gamma)(\sigma_{L,2}^{y}\sigma_{L+1,3}^{y}+\sigma_{L,2}^{y}\sigma_{L+1,4}^{y}+\sigma_{L,2}^{y}\sigma_{L+2,5}^{y}\\
&+\sigma_{L,3}^{y}\sigma_{L+2,4}^{y}+\sigma_{L,3}^{y}\sigma_{L+2,5}^{y}+\sigma_{L,4}^{y}\sigma_{L+3,5}^{y}).
\end{split}
\end{equation*}
This particular decomposition of the total Hamiltonian $H_{XY}^{(2d)}$ is able to guarantee the self-similarity of the renormalized Hamiltonian after each iterative step~\cite{18}.

The degenerate ground states for $H^{(2d)}_{B,L}$ are given by
\begin{equation}\label{eq:eq4}
\begin{split}
|\Upsilon_{0}\rangle&=\zeta_{1}(|\uparrow\uparrow\uparrow\uparrow\downarrow\rangle+|\uparrow\uparrow\uparrow\downarrow\uparrow\rangle+|\uparrow\uparrow\downarrow\uparrow\uparrow\rangle+|\uparrow\downarrow\uparrow\uparrow\uparrow\rangle)\\
&+\zeta_{2}(|\uparrow\uparrow\downarrow\downarrow\downarrow\rangle+|\uparrow\downarrow\uparrow\downarrow\downarrow\rangle+|\uparrow\downarrow\downarrow\uparrow\downarrow\rangle+|\uparrow\downarrow\downarrow\downarrow\uparrow\rangle)\\
&+\zeta_{3}|\downarrow\uparrow\uparrow\uparrow\uparrow\rangle+\zeta_{4}(|\downarrow\uparrow\uparrow\downarrow\downarrow\rangle+|\downarrow\uparrow\downarrow\uparrow\downarrow\rangle+|\downarrow\uparrow\downarrow\downarrow\uparrow\rangle\\
&+|\downarrow\downarrow\uparrow\uparrow\downarrow\rangle+|\downarrow\downarrow\uparrow\downarrow\uparrow\rangle+|\downarrow\downarrow\downarrow\uparrow\uparrow\rangle)+\zeta_{5}|\downarrow\downarrow\downarrow\downarrow\downarrow\rangle,
\end{split}
\end{equation}
and
\begin{equation}\label{eq:eq5}
\begin{split}
|\Upsilon_{1}\rangle&=\zeta_{6}|\uparrow\uparrow\uparrow\uparrow\uparrow\rangle+\zeta_{7}(|\uparrow\uparrow\uparrow\downarrow\downarrow\rangle+|\uparrow\uparrow\downarrow\uparrow\downarrow\rangle+|\uparrow\uparrow\downarrow\downarrow\uparrow\rangle\\
&+|\uparrow\downarrow\uparrow\uparrow\downarrow\rangle+|\uparrow\downarrow\uparrow\downarrow\uparrow\rangle+|\uparrow\downarrow\downarrow\uparrow\uparrow\rangle+\zeta_{8}|\uparrow\downarrow\downarrow\downarrow\downarrow\rangle\\
&+\zeta_{9}(|\downarrow\uparrow\uparrow\uparrow\downarrow\rangle+|\downarrow\uparrow\uparrow\downarrow\uparrow\rangle+|\downarrow\uparrow\downarrow\uparrow\uparrow\rangle+|\downarrow\downarrow\uparrow\uparrow\uparrow\rangle)\\
&+\zeta_{10}(|\downarrow\uparrow\downarrow\downarrow\downarrow\rangle+|\downarrow\downarrow\uparrow\downarrow\downarrow\rangle+|\downarrow\downarrow\downarrow\uparrow\downarrow\rangle+|\downarrow\downarrow\downarrow\downarrow\uparrow\rangle),
\end{split}
\end{equation}
where
\begin{equation*}
\zeta_{1}=-\frac{\sqrt{\varsigma+\gamma^{2}-1}}{4\sqrt{\varsigma}};~~\zeta_{2}=-\frac{\sqrt{\varsigma-\gamma^{2}+1}}{4\sqrt{\varsigma}}\Theta(\gamma);
\end{equation*}
\begin{equation*}
\zeta_{3}=\frac{\gamma^{2}-1+\varsigma}{\sqrt{4+6\gamma^{6}-2\eta_{1}}};~\zeta_{4}=\frac{\gamma(5+\gamma^{2}+\varsigma)}{2\sqrt{4+6\gamma^{6}-2\eta_{1}}};
\end{equation*}
\begin{equation*}
\zeta_{5}=\frac{3\sqrt{2}\gamma^{2}}{\sqrt{2+3\gamma^{6}-\eta_{1}}};~\zeta_{6}=\frac{|\gamma|(\gamma^{2}-1-\varsigma)}{\sqrt{4\gamma^{6}+2\eta_{2}}};
\end{equation*}
\begin{equation*}
\zeta_{7}=-\frac{\Theta(\gamma)(1+5\gamma^{2}+\varsigma)}{2\sqrt{4\gamma^{6}+2\eta_{2}}};~\zeta_{8}=-\frac{3\sqrt{2}|\gamma|}{\sqrt{2\gamma^{6}+\eta_{2}}};
\end{equation*}
\begin{equation*}
\zeta_{9}=\frac{\gamma}{4}\sqrt{\frac{1-\gamma^{2}+\varsigma}{\gamma^{2}\varsigma}};~~\zeta_{10}=\frac{1}{4}\sqrt{\frac{\gamma^{2}+\varsigma-1}{\varsigma}},
\end{equation*}
where $\eta_{1}=2\varsigma-\gamma^{4}(104+3\varsigma)-\gamma^{2}(71+17\varsigma)$, $\eta_{2}=\gamma^{4}(71-2\varsigma)+\gamma^{2}(104+17\varsigma)+3(\varsigma+1)$, $\varsigma=\sqrt{\gamma^{4}+34\gamma^{2}+1}$ for the case $\gamma\neq 0$ and the function $\Theta(\gamma)$ gives -1 or 1 depending on whether $\gamma$ is negative or positive. For the case $\gamma=0$, $\zeta_{2}=-\zeta_{9}=-\sqrt{2}/4$, $\zeta_{4}=-\zeta_{7}=\sqrt{3}/6$ and the others $\zeta_{i}=0$. The projection operator $\mathcal{P}^{(2d)}=\prod_{L}\mathcal{P}^{(2d)}_{L}$ can be constructed as $\mathcal{P}^{(2d)}_{L}=|\Uparrow\rangle\langle\Upsilon_{0}|+|\Downarrow\rangle\langle\Upsilon_{1}|$. Then, the effective (renormalized) Hamiltonian can be obtained as $\tilde{H}_{XY}^{(2d)}=\mathcal{P}^{(2d)\dagger}H_{XY}^{(2d)}\mathcal{P}^{(2d)}$ and the explicit expression is given by~\cite{18}
\begin{equation}\label{eq:eq6}
\begin{split}
\tilde{H}_{XY}^{(2d)}&=\frac{\acute{\lambda}}{4}\sum_{p=1}^{\mathcal{N}/5}\sum_{q=1}^{\mathcal{N}/5}[(1+\acute{\gamma})(\sigma_{p,q}^{x}\sigma_{p+1,q}^{x}+\sigma_{p,q}^{x}\sigma_{p,q+1}^{x})\\
&+(1-\acute{\gamma})(\sigma_{p,q}^{y}\sigma_{p+1,q}^{y}+\sigma_{p,q}^{y}\sigma_{p,q+1}^{y})],
\end{split}
\end{equation}
where $\acute{\lambda}=6\lambda\xi_{0}$ and $\acute{\gamma}=(2\xi_{1}+\gamma\xi_{2})/\xi_{0}$ with
\begin{equation*}
\begin{split}
\xi_{1}=&(3\zeta_{4}\zeta_{10}+3\zeta_{1}\zeta_{7}+\zeta_{2}\zeta_{8}+\zeta_{3}\zeta_{9})\\
&\times(\zeta_{5}\zeta_{10}+\zeta_{1}\zeta_{6}+3\zeta_{2}\zeta_{7}+3\zeta_{4}\zeta_{9}),
\end{split}
\end{equation*}
\begin{equation*}
\begin{split}
\xi_{2}&=\zeta_{10}^{2}(9\zeta_{4}^{2}+\zeta_{5}^{2})+\zeta_{1}^{2}(\zeta_{6}^{2}+9\zeta_{7}^{2})+\zeta_{2}^{2}(9\zeta_{7}^{2}+\zeta_{8}^{2})\\
&+2\zeta_{2}\zeta_{9}(9\zeta_{4}\zeta_{7}+\zeta_{3}\zeta_{8})+\zeta_{9}^{2}(\zeta_{3}^{2}+9\zeta_{4}^{2})+2\zeta_{10}\\
&\times(\zeta_{1}\zeta_{5}\zeta_{6}+9\zeta_{1}\zeta_{4}\zeta_{7}+3\zeta_{2}\zeta_{5}\zeta_{7}+3\zeta_{2}\zeta_{4}\zeta_{8}+3\zeta_{3}\zeta_{4}\zeta_{9}\\
&+3\zeta_{4}\zeta_{5}\zeta_{9})+6\zeta_{1}(\zeta_{2}\zeta_{6}\zeta_{7}+\zeta_{2}\zeta_{7}\zeta_{8}+\zeta_{4}\zeta_{6}\zeta_{9}+\zeta_{3}\zeta_{7}\zeta_{9}),
\end{split}
\end{equation*}
\begin{equation*}
\begin{split}
\xi_{0}&=\zeta_{10}^{2}(9\zeta_{4}^{2}+6\gamma\zeta_{4}\zeta_{5}+\zeta_{5}^{2})+\zeta_{1}^{2}(\zeta_{6}^{2}+6\gamma\zeta_{6}\zeta_{7}+9\zeta_{7}^{2})\\
&+\zeta_{2}^{2}(9\zeta_{7}^{2}+6\gamma\zeta_{7}\zeta_{8}+\zeta_{8}^{2})+2\zeta_{2}\zeta_{9}(3\gamma\zeta_{3}\zeta_{7}+9\zeta_{4}\zeta_{7}\\
&+\zeta_{3}\zeta_{8}+3\gamma\zeta_{4}\zeta_{8})+\zeta_{9}^{2}(\zeta_{3}^{2}+6\gamma\zeta_{3}\zeta_{4}+9\zeta_{4}^{2})+2\zeta_{1}[\zeta_{2}\\
&\times(3\zeta_{6}\zeta_{7}+9\gamma\zeta_{7}^{2}+\gamma\zeta_{6}\zeta_{8}+3\zeta_{7}\zeta_{8})+\zeta_{9}(\gamma\zeta_{3}\zeta_{6}+3\zeta_{4}\zeta_{6}\\
&+3\zeta_{3}\zeta_{7}+9\gamma\zeta_{4}\zeta_{7})]+2\zeta_{10}[\zeta_{1}\zeta_{5}\zeta_{6}+9\zeta_{1}\zeta_{4}\zeta_{7}+3\zeta_{2}\zeta_{5}\zeta_{7}\\
&+3\zeta_{2}\zeta_{4}\zeta_{8}+3\zeta_{3}\zeta_{4}\zeta_{9}+3\zeta_{4}\zeta_{5}\zeta_{9}+\gamma(3\zeta_{1}\zeta_{4}\zeta_{6}+9\zeta_{2}\zeta_{4}\zeta_{7}\\
&+3\zeta_{1}\zeta_{5}\zeta_{7}+\zeta_{2}\zeta_{5}\zeta_{8}+9\zeta_{4}^{2}\zeta_{9}+\zeta_{3}\zeta_{5}\zeta_{9})].
\end{split}
\end{equation*}
\begin{figure}
\centering
\includegraphics[angle=0,width=8.5cm]{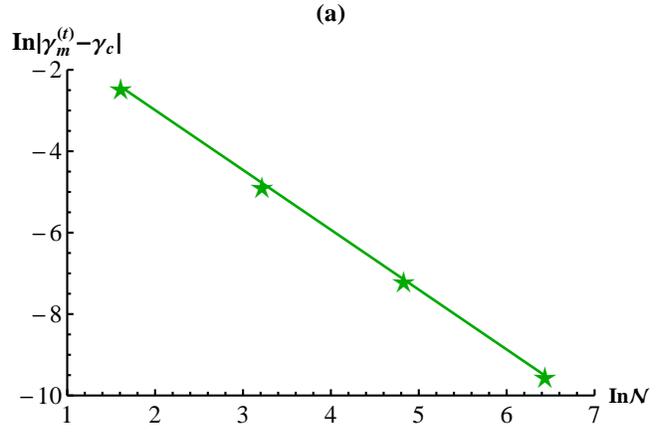}
\includegraphics[angle=0,width=8.5cm]{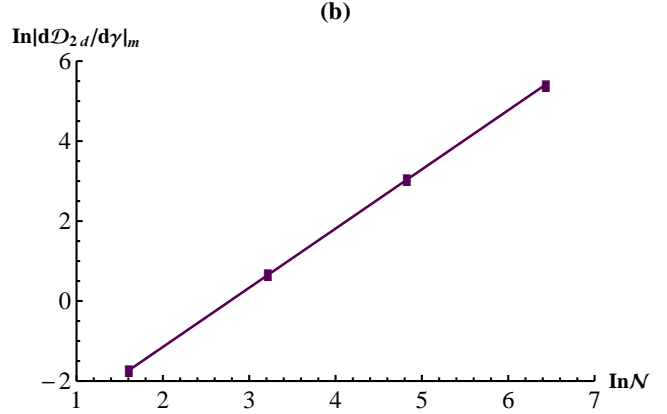}
\caption{\label{fig:fig7} (a) The logarithm of $|\gamma_{m}^{(t)}-\gamma_{c}|$ versus $\ln\mathcal{N}$. The green solid line is obtained by using the least-squares fit of the form: $\ln|\gamma_{m}^{(t)}-\gamma_{c}|=-\theta_{3}\ln\mathcal{N}+c_{3}$ where $\theta_{3}\simeq 1.470$ and $c_{3}\simeq 0.050$. (b) The logarithm of the maximum of $|d\mathcal{D}_{2d}/d\gamma|$ versus $\ln\mathcal{N}$. The purple solid line represents $\ln|d\mathcal{D}_{2d}/d\gamma|_{m}=\theta_{4}\ln\mathcal{N}+c_{4}$ where $\theta_{4}\simeq 1.475$ and $c_{4}\simeq-4.092$.}
\end{figure}

\begin{figure}
\centering
\includegraphics[angle=0,width=8.5cm]{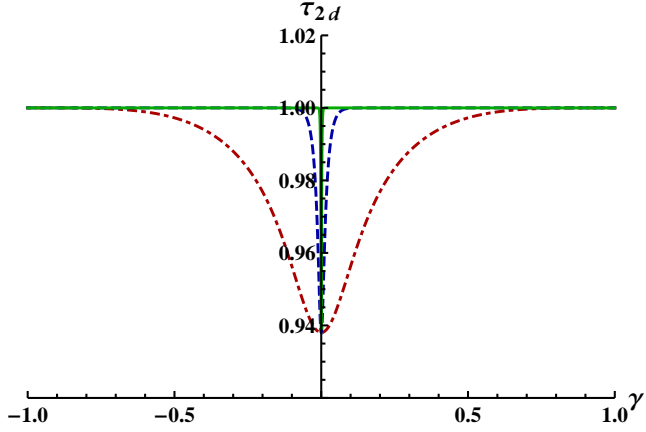}
\caption{\label{fig:fig8} The evolution of five-partite entanglement indicator $\tau_{2d}$ versus $\gamma$ in terms of QRG iteration steps: 0-th QRG step (red dotdashed line), 1-st QRG step (blue dashed line) and 2-nd QRG step (green solid line).}
\end{figure}

By solving the QRG equation $\acute{\gamma}=\gamma$, we find three solutions $\gamma=0,\pm1$ which is similar to the case of one-dimensional $XY$ spin-chain. Depending on the solutions, the model $H_{XY}^{(2d)}$ corresponds to a spin fluid phase for $\gamma\rightarrow0$ and an Ising-like phase for $\gamma\rightarrow\pm 1$. The QPT point obtained by QRG method is $\gamma_{c}=0$ which coincides with previous result~\cite{18,add1}.

Next, we study the trace distance of ground state $\varrho_{12345}=|\Upsilon_{0}\rangle\langle\Upsilon_{0}|$ and the Kronecker product of its marginal state $\varrho_{1234}\otimes\varrho_{5}$ which is given by $\mathcal{D}_{2d}\equiv\frac{1}{2}\||\Upsilon_{0}\rangle\langle\Upsilon_{0}|-\varrho_{1234}\otimes\varrho_{5}\|_{1}$,
where $\varrho_{1234}=tr_{5}(\varrho_{12345})$ and $\varrho_{5}=tr_{1234}(\varrho_{12345})$. It is worth pointing out that choosing $\varrho_{12345}=|\Upsilon_{1}\rangle\langle\Upsilon_{1}|$ yields the same physical results. We display the trace distance $\mathcal{D}_{2d}$ as the function of $\gamma$ for different QRG iteration steps in Fig.~\ref{fig:fig6}. It is found that the trace distance $\mathcal{D}_{2d}$ exhibits a sudden drop at the QPT point $\gamma=0$ and attains two fixed values: $\mathcal{D}_{2d}(\gamma=0)\simeq 0.718$ and $\mathcal{D}_{2d}(\gamma\neq0)=0.750$ after two QRG iteration steps. We also explore the finite-size scaling behaviors of renormalized trace distance $\mathcal{D}_{2d}$ in the vicinity of the quantum critical points in Fig.~\ref{fig:fig7}. It is shown that the pseudo-critical point $\gamma_{m}^{(t)}$ which denotes the position of the maximum of $|d\mathcal{D}_{2d}/d\gamma|$, approaches the real QPT point $\gamma_{c}$ exponentially with the increase of QRG iteration step, i.e., $\ln|\gamma_{m}^{(t)}-\gamma_{c}|\simeq-\theta_{3}\ln\mathcal{N}+const.$ (see Fig.~\ref{fig:fig7}(a)). This scaling behavior suggests that the
pseudo-critical points coincide with the real QPT points in the thermodynamics limit. The maximum of $|d\mathcal{D}_{2d}/d\gamma|$ obeys $\ln|d\mathcal{D}_{2d}/d\gamma|_{m}\simeq\theta_{4}\ln\mathcal{N}+const.$ (see Fig.~\ref{fig:fig7}(b)).

\begin{figure}
\centering
\includegraphics[angle=0,width=8.5cm]{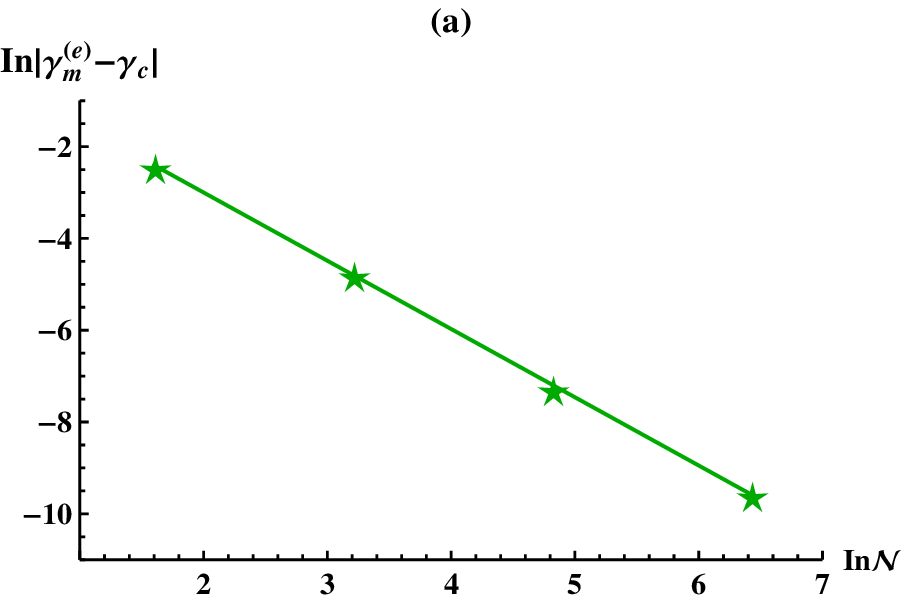}
\includegraphics[angle=0,width=8.5cm]{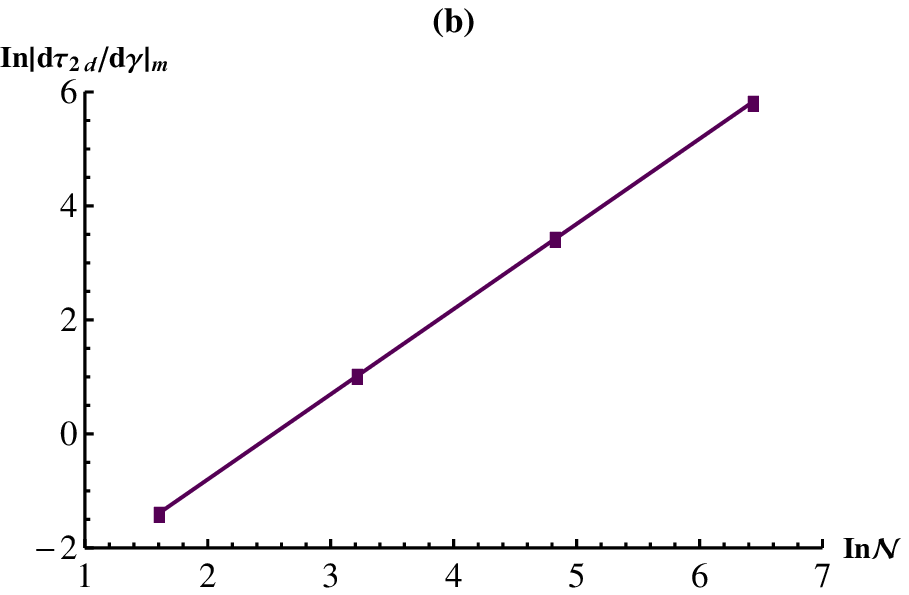}
\caption{\label{fig:fig9} (a)The logarithm of $|\gamma_{m}^{(e)}-\gamma_{c}|$ versus $\mathcal{N}$. The green solid line is obtained by using the least-squares fit of the form: $\ln|\gamma_{m}^{(e)}-\gamma_{c}|=-\theta_{5}\ln\mathcal{N}+c_{5}$ where $\theta_{5}\simeq 1.487$ and $c_{5}\simeq -0.025$. (b) The logarithm of the maximum of $|d\tau_{2d}/d\gamma|$ versus $\mathcal{N}$. The purple solid line represents $\ln|d\tau_{2d}/d\gamma|_{m}=\theta_{6}\ln\mathcal{N}+c_{6}$ where $\theta_{6}\simeq 1.494$ and $c_{6}\simeq-3.786$.}
\end{figure}

One can obtain a five-partite entanglement indicator which is defined by $\tau_{2d}\equiv E_{f}^{2}(\varrho^{1}_{1\mid 2345})-\sum_{k=2}^{5}E_{f}^{2}(\varrho_{1k})$. We plot the evolution of five-partite entanglement indicator $\tau_{2d}$ versus $\gamma$ with increasing the spin-size in Fig.~\ref{fig:fig8}. It is clearly seen that the five-partite entanglement $\tau_{2d}$ develops two rather different features after two QRG iteration steps. In the spin-fluid phase $\gamma\rightarrow0$, the the minimum value of five-partite entanglement is obtained, when $\gamma\neq 0$, the five-partite entanglement $\tau_{2d}$ reaches its maximum. It is shown that the pseudo-critical point $\gamma_{m}^{(e)}$, which corresponds to the position of the maximum of $|d\tau_{2d}/d\gamma|$, exhibits the scaling law: $\ln|\gamma_{m}^{(e)}-\gamma_{c}|\simeq-\theta_{5}\ln\mathcal{N}+const.$ (see Fig.~\ref{fig:fig9}(a)). We also find the maximum of $|d\tau_{2d}/d\gamma|$ obeys $\ln|d\tau_{2d}/d\gamma|_{m}\simeq\theta_{6}\ln\mathcal{N}+const$ (see Fig.~\ref{fig:fig9}(b)). Our numerical analysis shows $\theta_{3}\simeq\theta_{4}\simeq\theta_{5}\simeq\theta_{6}$ which is similar to that of spin-chains in one dimension. This result agrees with the expected universality and completes the analysis of trace distance and multipartite entanglement in two-dimensional $XY$ spin-chain.

We would like to compare our results with some standard known results. Quantum fidelity or fidelity susceptibility is one of the most widely used indicators of QPT in many-body systems, it is shown that the fidelity or fidelity susceptibility is able to exhibit a sudden jump near the QPT point~\cite{13,add2,add3}, this result indicates the derivative of fidelity or fidelity susceptibility is discontinuous at the QPT points in the thermodynamics limit which is similar to the behaviors of renormalized trace distance and multipartite entanglement employed in this paper. The QPT points revealed by our method coincide with standard known results~\cite{1,6,8,18,add1,add2}. Moreover, the correlation length exponent obtained by the QRG approach is $\nu=\theta^{-1}_{1,2}\simeq 1$ for the one-dimensional $XY$ chain, this result is also in good agreement with previous studies~\cite{add2,add3}.

In summary, we study the QPTs of a one- and a two-dimensional $XY$ spin-chains by employing trace distance and multipartite entanglement along with QRG method. It is found that the first derivatives of the renormalized trace distance and multipartite entanglement experience certain singular behaviors close to the quantum critical points of these spin-chain systems in the thermodynamics limit. This singular behaviors, which are closely associated with the divergence of the correlation length, become pronounced as the QRG iteration step increases. Furthermore, we find the renormalized trace distance and multipartite entanglement obey some universal and exponential-type scaling laws as the system size grows in the vicinity of the QPT points. The cornerstone of the theory of critical phenomena is the concept of universality, which indicates that the critical exponents are independent of the chosen order parameter~\cite{1,3,23}. Our results show that the critical exponents of trace distance and multipartite entanglement are identical in the same spin system, i.e., $\theta_{1}\simeq\theta_{2}$ for $XY$ spin chain in one dimension and $\theta_{3}\simeq\theta_{4}\simeq\theta_{5}\simeq\theta_{6}$ for $XY$ spin chain in two dimensions. This is in good agreement with the universality hypothesis of the quantum critical system. Our work may be reexamined by making use of numerical density matrix renormalization group (DMRG) method. The results presented in this paper can be extended to other many-body systems and may have some potential applications in quantum information science.

\acknowledgments

This project was supported by the National Natural Science Foundation of China (Grant No. 11274274) and the Fundamental Research Funds for the Central
Universities (Grant No. 2016FZA3004).



\end{document}